\documentclass[amsmath,amssymb,prb,twocolumn,superscriptaddress]{revtex4}

\usepackage{color}
\usepackage{graphicx}

\begin{document}

\title{Linear optical response\\ of current-carrying molecular junction:\\ 
A NEGF-TDDFT approach}
\date{\today}
\author{Michael Galperin}
\email{galperin@lanl.gov}
\author{Sergei Tretiak}
\affiliation{Theoretical Division, Center for Integrated Nanotechnologies,
             Los Alamos National Laboratory, Los Alamos, NM 87545}

\begin{abstract}
We propose a scheme for calculation of linear optical response of
current-carrying molecular junctions for the case when electronic tunneling
through the junction is much faster than characteristic time of 
external laser field.
We discuss relationships between nonequilibrium Green function (NEGF) and 
time-dependent density functional theory (TDDFT) approaches, and
derive expressions for optical response and linear polarizability within 
NEGF-TDDFT scheme. Corresponding results for isolated molecule,
derived within TDDFT approach previously, are reproduced when coupling
to contacts is neglected.
\end{abstract}

\maketitle

\section{\label{intro}Introduction}
Rapid progress of experimental capabilities in the field of molecular 
electronics necessiates development of theoretical (and calculational) tools
capable of explaining existing data and predicting (proposing) future
experiments. While initially focus of both experimental and theoretical studies 
was on ballistic current-voltage characteristic of molecular 
junctions,\cite{NitzanRatner}
today more complicated phenomena are on the forefront of research.
This includes inelastic transport (inelastic electron tunneling spectroscopy 
both far off\cite{StipeRezaeiHo_PRL} 
and at resonance\cite{Zhitenev}), 
current-induced motion,\cite{StipeRezaeiHo}
nonlinear conductance 
(negative differential resistance,\cite{Reed} 
hysteresis,\cite{Tour} 
and switching\cite{Blum}), 
shot noise,\cite{Ruitenbeek} 
Coulomb blockade,\cite{Bjornholm} 
Kondo effect,\cite{ParkPasupathy,Natelson} 
heating of molecular junctions,\cite{Nitzan,Cahill} etc. 
Recently optical properties of molecular 
junctions started to attract attention of researchers. Application of
external laser field promises ability to effectively control transport 
properties of molecular devices,\cite{Hanggi} 
while Raman spectroscopy, together with 
scanning tunneling microscopy (STM) and 
inelastic electron tunneling spectroscopy (IETS)  can serve as a 
diagnostic tool.\cite{Tao,Natelson2007}  First experimental data
in this direction include light induced switching behavior of molecular
junctions,\cite{DulicMolen,WakayamaOgawa,YasutomiMorita} 
voltage effects on fluorescence\cite{ParkBarbara} of molecules in 
nanojunctions, and surface enhanced Raman scattering (SERS) from molecules 
positioned in narrow gaps between metal 
nanoparticles.\cite{Moskovitz,NieEmory,KneipWang,JiangBosnik}   
Theoretically opto-electronic properties of current carrying molecular 
junctions have been studied mostly within simple 
models.\cite{BukerKirczenow,HOMOLUMO,HarbolaMaddoxMukamel,Cuevas} 
Here we make a first step in direction of ab initio calculations 
of optical properties of such junctions.

% NEGF in transport
Transport properties of molecular junctions are naturally described
within non-equilibrium Green's function (NEGF) 
approach.\cite{KadanoffBaym,Keldysh} 
A consistent way to treat transport of molecular junction 
within NEGF implies using many-body Green functions approach.
However, complexity of the schemes limits applicability of this treatment
to relatively simple molecular models 
only.\cite{Caroli,Datta,Ueba,GalperinNitzanRatner} 
% DFT isolated systems
DFT is a well-established tool for accurate calculation of ground state
electronic properties of isolated systems 
(atoms, molecules, solids).\cite{HohenbergKohn,KohnSham,ParrYang,Burke} 
% NEGF-DFT in transport
DFT ability to deal with relatively extended realistic systems
made the idea of merging the two approaches very appealing.
First NEGF-DFT was implemented to steady-state ballistic (Landauer) transport 
calculations.\cite{XueDattaRatner,DamleGhoshDatta,Transiesta} 
Later the calculations were extended to the case of inelastic electron 
transport in the weak electron-phonon coupling 
limit.\cite{LorentePersson,LorenteJauho,Pecchia,ReimersHush,SergueevDemkovGuo}
For detailed discussion on different approaches to transport 
through molecular junctions see Ref.~\onlinecite{review}.
While NEGF-DFT calculations modedling IETS reported good agreement 
with experimental data,\cite{LorenteJauho,Pecchia}
results for ballistic current-voltage calculations are not 
always in quantitative agreement.
Sometimes experimental data are reproduced by NEGF-DFT 
approach,\cite{Guo2006,Guo2007}
while in other cases calculations yield  
a current that is orders of magnitude larger than 
experimental values.\cite{Stokbro,Louie2007}

% Critique of NEGF-DFT
Besides obvious uncertainties inherent for molecular junction simulations,
such as influence of contact geometry\cite{BratkovskyKornilovich,Nara} 
and local environment\cite{SelzerTourAllara} on transport, 
a methodological DFT problem (when applied to transport situation) 
was pointed as a possible source of
discrepancy.\cite{EversWeigendKoentopp,Burke2005,KoentoppBurkeEvers,EversBurke}
Possible errors can be grouped into several categories:
1.~use of inappropriate exchange-correlation functional 
(ground state and/or spatially local character of functionals used,
self-interaction errors, absence of derivative discontinuity
and $xc$ contribution to the electric field response);
2.~inherent time-dependent character of transport, which
goes beyond validity of DFT;   
3.~DFT (as well as TDDFT discussed below) is a theory which work for 
finite systems (external disturbance local in space) only; 
4.~unphysical separation of the system into 
disconnected parts (contacts and molecule) at infinite past
within NEGF. 
For a detailed discussion on application of DFT in transport calculations see 
Refs.~\onlinecite{KoentoppBurkeCar,GebauerBurkeCar,GrossRubioLeeuwen}.

% Optics
Optical response of system is defined by its electronic excitations. 
In order to calculate excited states two main approaches can be implemented. 
One is based on a many-body theory, where solution of the Bethe-Salpeter 
equation (BSE) is needed. Its computationally expensive character limits
applicability of the approach to a relatively narrow range of 
problems (see e.g. Refs.~\onlinecite{Louie1999,Horst,Ruini}). 
A much more numerically efficient
alternative is the time dependent density-functional theory (TDDFT).
For a review comparing these two techniques see 
Ref.~\onlinecite{OnidaReiningRubio}. 
TDDFT is an extention of density functional theory,\cite{RungeGross,GrossKohn} 
which properly treats correlated excited states and
where electronic excitations are associated with the poles of exact 
charge density response. TDDFT is known for successful calculations 
of optical spectra in many
finite molecular systems\cite{OnidaReiningRubio,Casida2000} 
Its numerical simplicity allows treating systems that involve hundreds of atoms,
and a lot of work within the approach has been done for isolated 
molecules.\cite{ChernyakMukamel1995,ChernyakMukamel2000,TretiakChernyak2003,TretiakChernyak2005,TretiakKilina2007,Tretiak2007,PiryatshinskiTretiakCherniak2007,Maitra}
Taking into account size of realistic systems (molecules) used
in molecular devices, TDDFT is the only tool available today capable
of dealing with such calculations.

% TDDFT-NEGF in transport
Transport calculation schemes based on TDDFT were proposed as
an alternative to NEGF-DFT method.
Main differences are time-dependent character of
the TDDFT scheme (TDDFT instead of DFT), absence of system partitioning, 
and finite size of the system under study.
In particular, one of the proposed schemes, time dependent Kohn-Sham
master equation approach, utilizes ring geometry, and linearly
increasing  magnetic field in the center of the ring provides driving 
force.\cite{KoentoppBurkeCar,GebauerBurkeCar}  
Another TDDFT-NEGF approach considers time evolution of finite linear
system initially at equilibrium under influence of external field
(battery discharge).\cite{GrossRubioLeeuwen,StefanucciAlmbladh_EL,StefanucciAlmbladh_PRB,StefanucciAlmbladhGross}
While these approaches seem to solve (at least partially) problems of NEGF-DFT,
some questions still exist. For example, in the master equation approach,
finite size of the ring forces to apply artificially large coupling
between electrons and the bath in order to achieve thermalization
in the part of the ring representing contact, which rises question of
physicality of charge distribution in the contacts and, hence, current
through the device part of the ring. TDDFT-NEGF also implements
system of finite size, i.e. continuum character of states in the
electrodes is questionable. Thus equivalence of transient current 
calculated within the approach and realistic steady-state current
is not obvious. The approaches are still have to be tested
on the problems where NEGF-DFT failed. Note also that TDDFT
has its own limitations (nonuniqueness of the excited-state potentials,
question of stability and chaos of the mapping of densities on 
potentials).\cite{GaudoinBurke,Baer}

% Discussion on NEGF-DFT and NEGF-TDDFT relation in transport
While agreement on existence of methodological pitfalls of NEGF-DFT
exists, the importance of those errors (i.e. if they are the cause
of discrepancy between NEGF-DFT results and experimental data)
is not completely understood yet.\cite{EversBurke} 
Besides, within TDDFT-NEGF approach it was shown that after initial
correlations die out and steady-state current is established, the last is
given  by Landauer-like formula with chemical potentials in Fermi distribution
functions being shifted in accordance with extra exchange-correlation 
term originating from response to external field 
(bias).\cite{KoentoppBurkeCar,GrossRubioLeeuwen} 
The problems of NEGF-DFT and TDDFT-NEGF schemes seem to stem from essentially
different basic assumptions of the two (NEGF and DFT) theories, 
compatibility of those
is not clear. However this is the only practical tool available today capable
of dealing with realistic simulations of molecular junctions.

% Our goal here
In this paper we consider optical response of current carrying molecular 
junction within NEGF-TDDFT approach. We assume that initially steady-state
current across the junction is established, and then the system is
probed by an external laser field. The last is assumed to be a weak perturbation
on top of the non-equilibrium steady-state. The assumption works in the case
when timescale for electron transport is much shorter than the characteristic 
time of external field. 
In the opposite case (both times are comparable or the field
is quicker) one would need to consider time-dependent transport, either
within TDDFT-NEGF\cite{GrossRubioLeeuwen,StefanucciAlmbladh_EL,StefanucciAlmbladh_PRB,StefanucciAlmbladhGross}
or time-dependent NEGF\cite{JauhoWingreenMeir} approach, explicitly.
We postpone such consideration for future research.
In treating steady-state flux we assume that Landauer formula
is correct, while $xc$ and chemical potential are adjusted properly,
as is discussed in 
Refs.~\onlinecite{KoentoppBurkeCar,GrossRubioLeeuwen}  
So, the treatment formally looks like the standard NEGF-DFT, however
one has to keep in mind points mentioned above.

% Structure of the paper
In Section~\ref{model} we introduce model of molecular junction.
Section~\ref{negf2tddft} briefly discusses general differences between NEGF
and TDDFT approaches, and proposes a way to describe the latter in terms
of the former. In Section~\ref{current} we derive expressions for 
steady-state flux, while Section~\ref{optical} deals with optical response
of the current-carrying junction. Section~\ref{conclude} summarizes
our findings.

\section{\label{model}Model}
The model we employ consists of a molecule coupled to two contacts 
(left $L$ and right $R$).
Each contact is a reservoir of free charge carriers at its equilibrium,
i.e. characterized by its own electro-chemical potential $\mu_K$ ($K=L,R$).
We introduce second quantization field operators for electrons
in the molecule $\hat\psi_\sigma$ and in the contacts $\hat\psi_{\sigma,K}$ 
($\sigma$ is electron spin and $K=L,R$)
with the usual anti-commutation relations
\begin{align}
 \left\{\hat\psi_{\sigma_1}(\vec r_1;\mathbf{R});
        \hat\psi_{\sigma_2}^\dagger(\vec r_2;\mathbf{R})
 \right\} &= \delta_{\sigma_1,\sigma_2}\,\delta\left(\vec r_1-\vec r_2\right)
 \\
 \left\{\hat\psi_{\sigma_1,K_1}(\vec r_1);
        \hat\psi_{\sigma_2,K_2}^\dagger(\vec r_2)\right\}
 &= \delta_{\sigma_1,\sigma_2}\,\delta_{K_1,K_2}\,
    \delta\left(\vec r_1-\vec r_2\right)
\end{align}
and all other anti-commutators being zero. Note that molecular field operators
depend parametrically on the nuclear configuration 
$\mathbf{R}=\left\{\vec R_\alpha\right\}$ (Born-Oppenheimer approximation).

Many-body electronic Hamiltonian of the system in the second quantization is
\begin{equation}
 \label{H}
 \hat H(\mathbf{R}) = \hat H_L + \hat H_R + \hat H_M(\mathbf{R})
                    + \hat V_L(\mathbf{R}) + \hat V_R(\mathbf{R})
\end{equation}
where the contact Hamiltonian is given by
\begin{equation}
 \label{HK}
 \hat H_K = \sum_\sigma\int d\vec r_k\, \hat\psi_{\sigma,K}^\dagger(\vec r_k)
 \left[-\frac{\hbar^2}{2m}\frac{\partial^2}{\partial{\vec r}^2_k}\right]
 \hat\psi_{\sigma,K}(\vec r_k)
\end{equation}
(here and below $K=L,R$), 
coupling between molecule and contact $K$ is (no spin-flip transitions)
\begin{align}
 \label{VK}
 \hat V_K(\mathbf{R}) =& \sum_\sigma\int d\vec r\int d\vec r_k\,
 \\
 &\left[\hat\psi_\sigma^\dagger(\vec r;\mathbf{R})\,
 V_{\sigma,K}(\vec r,\vec r_k;\mathbf{R})\,\hat\psi_{\sigma,K}(\hat r_k) 
 + \mbox{H.c.}\right],
 \nonumber
\end{align}
and molecular Hamiltonian is
\begin{align}
 \label{HM}
 &\hat H_M(\mathbf{R}) = \sum_{\sigma}\int d\vec r\,
 \hat\psi_\sigma^\dagger(\vec r;\mathbf{R}) 
 \left[
 -\frac{\hbar^2}{2m}\frac{\partial^2}{\partial{\vec r}^2}
 \right.\nonumber\\ &\left. \qquad\qquad\qquad
 -\sum_\alpha\frac{Z_\alpha}{\left|\vec r-\vec R_\alpha\right|} 
 - e\vec E(t)\vec r
 \right] \hat\psi_\sigma(\vec r;\mathbf{R})
 \\
 &+ \frac{e^2}{2}\sum_{\sigma_1,\sigma_2}\int d\vec r_1\int d\vec r_2\,
 \hat\psi_{\sigma_1}^\dagger(\vec r_1;\mathbf{R})\,
 \hat\psi_{\sigma_2}^\dagger(\vec r_2;\mathbf{R})
 \nonumber\\ &\qquad\qquad\qquad\qquad\times
 \frac{1}{\left|\vec r_1-\vec r_2\right|}
 \hat\psi_{\sigma_2}(\vec r_2;\mathbf{R})\,
 \hat\psi_{\sigma_1}(\vec r_1;\mathbf{R}).
 \nonumber
\end{align}
Here $\alpha$ indicates sum over nuclei of the molecule,
$Z_\alpha$ is a charge of the nucleus $\alpha$, and $\vec E(t)$
is external laser field.
Note that we intentionally started from the Hamiltonian in the second 
quantization form 
and used explicit partition of the system into contacts and a molecule.
Since this is a standard NEGF approach, it makes the following connection 
to TDDFT clearer. At the same time, NEGF partitioning scheme by 
Caroli~et~al.\cite{Caroli} and partitionless approach by Cini\cite{Cini}
implemented in TDDFT-NEGF schemes for time-dependent transport\cite{KoentoppBurkeCar,GebauerBurkeCar,GrossRubioLeeuwen,StefanucciAlmbladh_EL,StefanucciAlmbladh_PRB,StefanucciAlmbladhGross} 
in the steady-state case, which we are going to consider, 
were shown to be equivalent.\cite{StefanucciAlmbladh_EL}
Note that below we will consistently use atomic units, i.e. $\hbar=e=m=1$,
and drop $\mathbf{R}$, keeping in mind that all the quantities
depend on positions of nuclei parametrically.

Now in the spirit of DFT (and TDDFT) theory we replace the true many-body
molecular Hamiltonian (\ref{HM}) by fictitious single-particle
Kohn-Sham Hamiltonian, which in the second quantization form is
\begin{align}
 \label{HM_KS}
 \hat H_M^{(KS)} =& \sum_{\sigma}\int d\vec r\,
 \hat\psi_{\sigma}^\dagger(\vec r)\left[
 -\frac{1}{2}\frac{\partial^2}{\partial{\vec r}^2}
 + v^{ext}(\vec r) + v_{\sigma}^{cl}(\vec r,t) 
 \right.\nonumber\\ &\left.\qquad\qquad\qquad
 + v_{\sigma}^{xc}(\vec r,t)
 - \vec E(t)\vec r
 \right]\hat\psi_{\sigma}(\vec r)
\end{align}
Here
\begin{equation}
 \label{vext}
 v^{ext}(\vec r)=-\sum_\alpha\frac{Z_\alpha}{\left|\vec r-\vec R_\alpha\right|}
\end{equation}
is external potential due to electron-nuclear interaction,
\begin{equation}
 \label{vcl}
 v_{\sigma}^{cl}(\vec r,t) = \int d\vec r_1\,
 \frac{n_\sigma(\vec r_1,t)}{\left|\vec r-\vec r_1\right|}
\end{equation} 
is electron-electron Coulomb interaction,
and 
\begin{equation}
 \label{vxc}
 v_{\sigma}^{xc}(\vec r,t) = \frac{\delta A^{xc}[n]}{\delta n_\sigma(\vec r,t)}
\end{equation}
is the exchange-correlation potential.
Here $A^{xc}[n]$ is the exchange-correlation action, which should be
defined on the Keldysh contour.\cite{Leeuwen1998} 
In Eqs.~(\ref{vcl}) and (\ref{vxc}) $n_\sigma(r,t)$ is an electron density 
of spin $\sigma$ in position $\vec r$ at time $t$.

Below we will represent molecular electron subspace in some finite basis set 
$\{\phi_{i\sigma}\}$, where $i$ is some quantum number 
(or set of quantum numbers) and $\sigma$ is spin.
As a basis one can use atomic or molecular single-particle states (orbitals),
or any other convenient basis set. In this basis set representation
\begin{align}
 \hat\psi_\sigma(\vec r) &= \sum_i \hat c_{i\sigma}\,\phi_{i\sigma}(\vec r)
 \\
 \hat\psi_\sigma^\dagger(\vec r) &= \sum_i \hat c^\dagger_{i\sigma}\,
 \phi_{i\sigma}^{*}(\vec r)
\end{align}
where $\hat c_{i\sigma}$ ($\hat c^\dagger_{i\sigma}$) is electron
annihilation (creation) operator for state $i\sigma$.
Kinetic energy and potentials in Eqs.~(\ref{HM_KS})-(\ref{vxc}) become 
matrices in Hilbert space.
Similar basis set representation will be used for electrons in contacts.

\section{\label{negf2tddft}Density matrix formulation of NEGF}
Central quantity of interest within NEGF is an electron Green function (GF),
defined on the Keldysh contour as
\begin{equation}
 \label{Gdef}
 G_{ij\sigma}(\tau,\tau') = -i<T_c\, \hat c_{i\sigma}(\tau)\,
                                     \hat c_{j\sigma}^\dagger(\tau')>
\end{equation}
where $\tau$ and $\tau'$ are contour variables, $T_c$ is a contour ordering
operator, and operators $c_{i\sigma}$ are given in 
the Heisenberg representation. 
Applying standard perturbation theory and assuming that initial
correlations died out, one arrives at the Dyson equation on the contour
\begin{equation}
 \label{Dyson_left}
 \left[i\frac{\overset{\rightarrow}{\partial}}{\partial\tau}-H\right] 
 G(\tau,\tau') = \delta(\tau,\tau') + \int_c d\tau_1\, 
 \Sigma(\tau,\tau_1)\,G(\tau_1,\tau') 
\end{equation}
or relative to the other variable
\begin{equation}
 \label{Dyson_right}
 G(\tau,\tau')
 \left[-i\frac{\overset{\leftarrow}{\partial}}{\partial\tau'}-H\right]
 = \delta(\tau,\tau') + \int_c d\tau_1\, G(\tau,\tau_1)\,\Sigma(\tau_1,\tau').
\end{equation}
In (\ref{Dyson_left}) and (\ref{Dyson_right}),
$H$ is Hamiltonian of the part of the system
under study, while $\Sigma$ is self-energy (SE) which represents influence
of all other parts (and processes). 
Here and below we suppress matrix indices keeping in mind that 
Hamiltonian, GFs, and SEs are matrices in the Hilbert space. 
In the model introduced in Section~\ref{model} $H$ corresponds to
$H_M$, Eq.(\ref{HM}), and $\Sigma$ represents contacts and coupling to them,
Eqs.~(\ref{HK}) and (\ref{VK})
\begin{equation}
 \label{SEdef}
 \Sigma_{ij\sigma}(\tau_1,\tau_2) = \sum_{K=L,R}\sum_{k\in K}
 V_{ik,\sigma}\, g_{k\sigma}(\tau_1,\tau_2)\, V_{kj,\sigma} 
\end{equation}
where $g_{k\sigma}(\tau_1,\tau_2)=-i<T_c\, \hat c_{k\sigma}(\tau_1)\,
\hat c^\dagger_{k\sigma}(\tau_2)>$ is GF for free electrons in the
contacts.  Generally $\Sigma$ should include also
contributions from many-body processes, such as electron-electron
interaction in Eq.(\ref{HM}), however in an attempt to combine NEGF
with TDDFT the last is introduced through potentials (\ref{vcl}) and 
(\ref{vxc}) in the Kohn-Sham Hamiltonian (\ref{HM_KS}).

In what follows we will need lesser projection of (\ref{Dyson_left}) and 
(\ref{Dyson_right})
\begin{align}
 \label{Glt_left}
 &\left[i\frac{\overset{\rightarrow}{\partial}}{\partial t}-H\right] 
 G^{<}(t,t') = 
 \\
 &\int_{-\infty}^{+\infty} dt_1\,\left[ 
  \Sigma^{r}(t,t_1)\,G^{<}(t_1,t') + \Sigma^{<}(t,t_1)\,G^{a}(t_1,t') 
 \right]
 \nonumber\\
 \label{Glt_right}
 &G^{<}(t,t')
 \left[-i\frac{\overset{\leftarrow}{\partial}}{\partial t'}-H\right] = 
 \\
 &\int_{-\infty}^{+\infty} dt_1\, \left[
 G^{<}(t,t_1)\,\Sigma^{a}(t_1,t') + G^{r}(t,t_1)\,\Sigma^{<}(t_1,t')
 \right]
 \nonumber
\end{align}
and retarded projection of (\ref{Dyson_left})
\begin{equation}
 \label{Gr_left}
 \left[i\frac{\overset{\rightarrow}{\partial}}{\partial t}-H\right] 
 G^r(t,t') = \delta(t-t') + \int_{-\infty}^{+\infty} dt_1\, 
 \Sigma^r(t,t_1)\,G^r(t_1,t') 
\end{equation}
while advanced GF is
\begin{equation}
 \label{Ga}
 G^a_{ij\sigma}(t,t') = \left[G^r_{ji\sigma}(t',t)\right]^{*}
\end{equation}
Note, $t$ and $t'$ in (\ref{Glt_left})-(\ref{Ga}) are time variables.
In order to get time-dependent solution (without initial correlations) 
within NEGF one has to solve (\ref{Glt_left}) and (\ref{Gr_left}) 
simultaneously, while SE should represent also electron-electron
interaction.

While in NEGF one deals with retarded and lesser
\begin{equation}
 \label{Gltdef}
 G^{<}_{ij\sigma}(t,t') = i<\hat c_{j\sigma}^\dagger(t')\,\hat c_{i\sigma}(t)>
\end{equation}
GFs, central object of TDDFT is single-electron density matrix (DM)
\begin{equation}
 \label{DM}
 \rho_{ij\sigma}(t) = <\hat c_{j\sigma}^\dagger(t)\,\hat c_{i\sigma}(t)>
 \equiv -iG^{<}_{ij\sigma}(t,t)
\end{equation}
Note that rigorous foundations of TDDFT formalism\cite{RungeGross} 
establish correspondense only between densities of real and noninteracting 
(Kohn-Sham) systems. However in practice non-diagonal elements of the 
Kohn-Sham DMs are also used in calculations of optical properties.   
Moreover, approaches explicitly utilizing DM within time-dependent 
functional theory (TDDMFT) were also
proposed.\cite{Furche,Gross2005,Baerends}

Time evolution of DM
\begin{equation}
 \frac{\partial}{\partial t} \rho(t) = 
 -i\left[\frac{\partial}{\partial t}G^{<}(t,t') +
         \frac{\partial}{\partial t'}G^{<}(t,t')\right]_{t=t'}
\end{equation}
can be obtained from Eqs.~(\ref{Glt_left}) and (\ref{Glt_right}) 
(approximately) expressed in terms of $G^{<}(t,t)$ only.
In order to do so we employ generalized Kadanoff-Baym ansatz 
(GKBA)\cite{HaugJauho} in the right side of (\ref{Glt_left}) 
and (\ref{Glt_right}), thus partly loosing non-locality in time.
This leads to
\begin{align}
 \label{Glt_tt}
 &i\frac{\partial}{\partial t}G^{<}(t,t) - [H;G^{<}(t,t)]
 \\
 &- \int_{-\infty}^{+\infty}dt_1\left[
   \Sigma^r(t,t_1)\,G^{<}(t_1,t_1) - G^{<}(t_1,t_1)\,\Sigma^{a}(t_1,t)\right]
 \nonumber \\
 &= \int_{-\infty}^{+\infty}dt_1\left[
   G^r(t,t_1)\,\Sigma^{<}(t_1,t) - \Sigma^{<}(t,t_1)\,G^a(t_1,t)\right]
 \nonumber
\end{align}
Note that in the absence of contacts (i.e. when all SEs are zero)
and substituting $H_M^{(KS)}$, Eq.(\ref{HM_KS}), in place of $H$
we recover the standard TDDFT formulation in terms of 
DM.\cite{ChernyakMukamel2000}
Eq.(\ref{Glt_tt}) together with (\ref{Gr_left}) and (\ref{Ga})
is the approximate formulation for NEGF in terms of DM evolution.
Below we use these expressions in order to get first
a steady-state transport through the junction, and then 
optical response of such current carrying junction to an external laser field.

Summarizing, NEGF-TDDFT is superior over NEGF-DFT due to ability of 
treating both time-dependent transport and/or
optical response of current-carrying molecular junctions. 
Still, as is indicated above, it misses nonlocality in time due to 
GKBA applied and keeps limitations of TDDFT. Also fundamental question of
combining the two ideologically different schemes remains.

\section{\label{current}Steady-state current}
First we consider steady-state current through the junction in the absence
of an external field, $\vec E(t)=0$. In the steady-state situation GFs and SEs
in (\ref{Gr_left}) and (\ref{Glt_tt}) depend on the time difference only,
and DM becomes time-independent, $\bar\rho=-i\bar G^{<}(t=0)$. 
In order to simplify notation we further consider contacts 
within wide-band approximation (WBA), when
\begin{equation}
 \label{wbl}
 \Sigma^r(t_1-t_2)\approx -\frac{i}{2}\Gamma \delta(t_1-t_2)
\end{equation}
Here
\begin{equation}
 \label{Gamma}
 \Gamma_{ij\sigma} = 2\pi \sum_{K=L,R}\sum_{k\in K} V_{ik\sigma} V_{kj\sigma}
 \delta(E-\varepsilon_{k\sigma})
 \equiv \Gamma^L_{ij\sigma} + \Gamma^R_{ij\sigma}
\end{equation}
is assumed to be constant independent of energy $E$, 
$\varepsilon_{k\sigma}$ is energy of the state $k\sigma$.

In this case Eq.(\ref{Glt_tt}) yields (for brevity we write $\bar G^{<}$
keeping in mind that this is $\bar G^{<}(t=0)$)
\begin{align}
 \label{ss_Glt}
 &i\frac{\partial}{\partial t}\bar G^{<} - [\bar H;\bar G^{<}]
 +\frac{i}{2}\left\{\Gamma;\bar G^{<}\right\} =
 \\
 &\int_{-\infty}^{+\infty}dt_1\left[
   \bar G^r(t-t_1)\,\Sigma^{<}(t_1-t) - 
   \Sigma^{<}(t-t_1)\,\bar G^a(t_1-t)\right]
 \nonumber
\end{align}
where $[\ldots;\ldots]$ is a commutator, while $\{\ldots;\ldots\}$ is
an anti-commutator.
Note, $i\frac{\partial}{\partial t}\bar G^{<}=0$ and is written here 
only in order to keep similarity to the structure to Eq.(\ref{Glt_tt}).
Here
\begin{equation}
 \label{ss_H}
 \bar H_{ij\sigma} = \left(H_M^{(KS)}[\bar\rho]\right)_{ij\sigma}
 = h_{ij\sigma} + \bar v^{cl}_{ij\sigma} + \bar v^{xc}_{ij\sigma}
\end{equation}
with
\begin{align}
 \label{h}
 h_{ij\sigma} &= \int d\vec r\phi_{i\sigma}^{*}(\vec r)\left[
 -\frac{1}{2}\frac{\partial^2}{\partial{\vec r}^2}-v^{ext}(\vec r)\right] 
 \phi_{j\sigma}(\vec r)
 \\
 \label{ss_vcl}
 \bar v^{cl}_{ij\sigma} &= \sum_{m,n\sigma'}(ij\sigma|nm\sigma')
 \bar \rho_{mn\sigma'}
 \\
 (ij\sigma|nm\sigma') &= \int d\vec r_1\int d\vec r_2
 \phi^{*}_{i\sigma}(\vec r_1)\,\phi_{j\sigma}(\vec r_1) 
 \\ &\qquad\qquad\times
 \frac{1}{\left|\vec r_1-\vec r_2\right|} 
  \phi^{*}_{n\sigma'}(\vec r_2)\,\phi_{m\sigma'}(\vec r_2) 
 \nonumber
\end{align}
Note that $\bar v^{xc}$ generally is not a ground state $xc$ functional.
In order to estimate it one can follow approach described in 
Ref.~\onlinecite{Leeuwen1996} and utilize formal equivalence 
(in particular cases) 
between partitionless TDDFT and partitioned NEGF schemes for
the case of established steady-state current.\cite{StefanucciAlmbladh_PRB}
For details of approximate way to estimate $\bar v^{xc}$ see Appendix~\ref{A}.

Fourier transform (FT) of lesser SE entering Eq.(\ref{ss_Glt}) is
\begin{equation}
 \label{SElt}
 \Sigma^{<}(E) = i\left[f_L(E)\Gamma^L + f_R(E)\Gamma^R\right]
\end{equation}
where $f_K(E)$ is the Fermi distribution in the contact $K$,
and retarded GF $\bar G^r$ can be obtained from FT of (\ref{Gr_left})
\begin{equation}
 \label{ss_Gr}
 \bar G^r(E) = \left[E-\bar H + i\Gamma/2\right]^{-1}
\end{equation} 
Note that chemical potentials in the contacts should be shifted 
to take into account $xc$ response to bias induced 
field.\cite{StefanucciAlmbladh_EL}

Integral version of Eq.(\ref{ss_Glt}) (Keldysh equation)
\begin{equation}
 \label{ss_Glt_integral}
 \bar G^{<}(t=0) = \int_{-\infty}^{+\infty}\frac{dE}{2\pi}\,
 \bar G^r(E)\,\Sigma^{<}(E)\,\bar G^a(E),
\end{equation}
together with Eq.(\ref{ss_Gr}) yields
the standard NEGF-DFT approach to steady-state transport. 
Note however that in the last case $\bar v_{xc}$ is
substituted by ground state $xc$ potential and
$xc$ corrections to chemical potentials of the contacts 
are neglected.

\section{\label{optical}Linear optical response}
Here we consider linear optical response to weak external 
laser field $\vec E(t)$.
In contrast to previous TDDFT 
considerations\cite{ChernyakMukamel2000,TretiakChernyak2003}
the response is calculated on top of non-equilibrium steady-state 
(rather than ground state) of the system.  

Upon introduction of the time-dependent external field Eqs.~(\ref{Glt_tt})
and (\ref{Gr_left}) yield (within WBA)
\begin{align}
 \label{e_Glt}
 &i\frac{\partial}{\partial t}G^{<}(t,t) - [H;G^{<}(t,t)]
 +\frac{i}{2}\left\{\Gamma;G^{<}(t,t)\right\} =
 \\
 &\int_{-\infty}^{+\infty}dt_1\left[
   G^r(t,t_1)\,\Sigma^{<}(t_1-t) - \Sigma^{<}(t-t_1)\,G^a(t_1,t)\right]
 \nonumber \\
 \label{e_Gr}
 & G^r(t,t') = \bar G^r(t-t')
 \\ &\qquad\qquad +\int_{-\infty}^{+\infty}dt_1\,
   \bar G^r(t-t_1) \left[H-\bar H\right] G^r(t_1,t')
 \nonumber
\end{align} 
Note, in Eq.(\ref{e_Glt}) we assume that contacts are not influenced
by an external field (SEs are the same), and Eq.(\ref{e_Gr}) is
an integral variant of Eq.(\ref{Gr_left}). Here
\begin{align}
 \label{e_H}
 H_{ij\sigma} &= \left(H_M^{(KS)}[\rho(t)]\right)_{ij\sigma}
 \nonumber \\
 &= h_{ij\sigma} + v^{cl}_{ij\sigma}(t) + v^{xc}_{ij\sigma}(t)
 - \vec \mu_{ij\sigma}\vec E(t)
\end{align}
with $h_{ij\sigma}$ defined in Eq.(\ref{h}),
$v^{cl}_{ij\sigma}(t)$ defined similar to Eq.(\ref{ss_vcl})
with $\bar\rho$ replaced by $\rho(t)$, 
$v^{xc}_{ij\sigma}(t)$ is $xc$ potential when both bias and external 
optical field are applied to the junction, 
and the matrix element of molecular dipole
\begin{equation}
 \label{mudef}
 \vec \mu_{ij\sigma} = \int d\vec r\,\phi_{i\sigma}^{*}(\vec r)\, \vec r\,
                       \phi_{j\sigma}(\vec r)
\end{equation}
$\bar H$ is defined in Eq.(\ref{ss_H}).

Assuming external optical field is a weak perturbation, we linearize
Eqs.~(\ref{e_Glt}) and (\ref{e_Gr}) in $\vec E(t)$. 
Introducing response to the field
\begin{align}
 \label{deltaGlt}
 \delta G^{<}(t,t) &= G^{<}(t,t) - \bar G^{<} \equiv i\,\delta\rho(t)
 \\
 \label{delta Gra}
 \delta G^{r,a}(t,t') &= G^{r,a}(t,t') - \bar G^{r,a}(t-t'),
\end{align}
linearizing Eqs.~(\ref{e_Glt}) and (\ref{e_Gr}) 
in the field and response to it,
and subtracting Eq.(\ref{ss_Glt}) from the linearized version of 
Eq.(\ref{e_Glt}) we obtain
\begin{align}
 \label{e_deltaGlt}
 &i\frac{\partial}{\partial t}\delta G^{<}(t,t) - [\bar H;\delta G^{<}(t,t)]
 +\frac{i}{2}\left\{\Gamma;\delta G^{<}(t,t)\right\}
 \nonumber \\
 &\qquad\quad\  - [\delta v^{cl}(t)+\delta v^{xc}(t);\bar G^{<}] 
 + \vec E(t) [\vec\mu;\bar G^{<}]=
 \\
 &\int_{-\infty}^{+\infty}dt_1\left[
    \delta G^r(t,t_1)\,\Sigma^{<}(t_1-t) - \Sigma^{<}(t-t_1)\,\delta G^a(t_1,t)
    \right]
 \nonumber \\
 \label{e_deltaGr}
 & \delta G^r(t,t') = \int_{-\infty}^{+\infty}dt_1\,
   \bar G^r(t-t_1) 
 \\ &\qquad\qquad\times
\left[\delta v^{cl}(t_1)+\delta v^{xc}(t_1)
   - \vec E(t_1)\,\vec\mu\right] \bar G^r(t_1-t')
 \nonumber
\end{align}
Here
\begin{align}
 \delta v^{cl}_{ij\sigma}(t) &= v^{cl}_{ij\sigma}(t) - \bar v^{cl}_{ij\sigma}
 = \sum_{m,n,\sigma'} (ij\sigma|nm\sigma')\,\delta\rho_{mn\sigma'}(t)
 \\
 \delta v^{xc}_{ij\sigma}(t) &= v^{xc}_{ij\sigma}(t) - \bar v^{xc}_{ij\sigma}
 \nonumber \\
 &\approx \sum_{m,n,\sigma'}\int_{-\infty}^{+\infty}dt_1\,
 f^r_{ij\sigma,nm\sigma'}(t-t_1)\,\delta\rho_{mn\sigma'}(t_1) 
\end{align}
where superscript ``r'' on $xc$ response function $f^r(t-t_1)$ indicates
its retarded character, i.e. for $t<t_1$ $f^r(t-t_1)=0$.
Since steady state can be (approximately) mapped to an effective 
equilibrium,\cite{Zubarev,Hershfield,Han} $\delta v^{xc}(t)$ in principle
can be estimated following procedure of Ref.~\onlinecite{Leeuwen1996} 
when equilibrium is understood as an effective equilibrium corresponding
to a particular steady-state.

Substituting Eq.(\ref{e_deltaGr}) and its advanced analog into 
Eq.(\ref{e_deltaGlt}) and rearranging terms so that all source terms
(terms containing external field) are in the right-hand side,
we obtain final result, which in Liouville space is given by
\begin{align}
 \label{e_deltaGlt_final}
 &\left[i\frac{\partial}{\partial t}-L_0\right]\delta\rho^{<}(t)
 - \int_{-\infty}^{+\infty}dt_1\, L_1^r(t-t_1)\,\delta\rho(t_1)
 \nonumber \\
 &\qquad = \int_{-\infty}^{+\infty}dt_1\, \vec{\cal E}^r(t-t_1)\,\vec E(t_1)
\end{align}
where superscript ``r'' again indicates retarded character of the 
corresponding functions. Fourier transform of Eq.(\ref{e_deltaGlt_final}) is
\begin{equation}
 \label{e_deltaGlt_final_FT}
 \left[\omega-L_0-L_1^r(\omega)\right]\delta\rho(\omega) =
 \vec{\cal E}^r(\omega)\,\vec E(\omega)
\end{equation} 
Here Markovian part of the Liouvillian is
\begin{align}
 \label{L0}
 \left[L_0\right]_{ij\sigma,mn\sigma'} =&
 \left\{
  \left[\bar H_{im\sigma}-\frac{i}{2}\Gamma_{im\sigma}\right]\delta_{nj}
 \right. \\ &\left. 
 -\left[\bar H_{nj\sigma}+\frac{i}{2}\Gamma_{nj\sigma}\right]\delta_{im}
 \right\}\delta_{\sigma\sigma'},
 \nonumber
\end{align}
frequency-dependent Liouvillian is
\begin{align}
 \label{L1}
 &\left[L_1^r(\omega)\right]_{ij\sigma,mn\sigma'} = 
 \nonumber \\
 &\sum_p \left[ f^r_{ip\sigma,nm\sigma'}(\omega)\,\bar\rho_{pj\sigma}-
                \bar\rho_{ip\sigma}\, f^r_{pj\sigma,nm\sigma'}(\omega)
         \right]
 \nonumber \\
 & -i\sum_{p,q,r}\int_{-\infty}^{+\infty}\frac{dE}{2\pi}
 \left\{ \bar G^r_{ip\sigma}(E+\omega)\,\bar G^r_{qr\sigma}(E)\,
         \Sigma^{<}_{rj\sigma}(E)
 \right.  \\ 
 &\qquad\qquad\qquad\qquad\times
 \left[(pq\sigma|nm\sigma') + f^r_{pq\sigma,nm\sigma'}(\omega)\right]
 \nonumber \\ &\qquad\qquad\qquad\quad
 -\Sigma^{<}_{ir\sigma}(E)\bar G^a_{rq\sigma}(E)\,\bar G^a_{pj\sigma}(E-\omega)
 \nonumber \\ &\qquad\qquad\qquad\qquad\left.\times
 \left[(qp\sigma|nm\sigma') + f^r_{qp\sigma,nm\sigma'}(\omega)\right]
 \right\},
 \nonumber
\end{align}
and frequency-dependent coefficient of the source term is
\begin{align}
 \label{calE}
 \vec{\cal E}^r_{ij}(\omega) =& 
  \sum_{p,\sigma}\left[\vec\mu_{ip\sigma}\,\bar\rho_{pj\sigma}
              -\bar\rho_{ip\sigma}\,\vec\mu_{pj\sigma}\right]
 -i\sum_{p,q,r,\sigma}\int_{-\infty}^{+\infty}\frac{dE}{2\pi}
 \nonumber \\
 &\ \ \left[\bar G^r_{ip\sigma}(E+\omega)\,\bar G^r_{qr\sigma}(E)\,
       \Sigma^{<}_{rj\sigma}(E)\,\vec\mu_{pq\sigma} 
 \right. \\ &- \left.
       \Sigma^{<}_{ir\sigma}(E)\,\bar G^a_{rq\sigma}(E)\,
       \bar G^a_{pj\sigma}(E-\omega)\,\vec\mu_{qp\sigma}
 \right].
 \nonumber
\end{align}

Since in the linear optical response only particle-hole, 
$\bar\rho_{ii\sigma}>\bar\rho_{jj\sigma}$, and hole-particle,
$\bar\rho_{ii\sigma}<\bar\rho_{jj\sigma}$, elements of
$\delta\rho_{ij\sigma}(t)$ are nonzero,\cite{Casida}
size of the matrix equation (\ref{e_deltaGlt_final_FT}) can be reduced.
Following Ref.~\onlinecite{Casida} we order the basis such that
$i<j$ for $\bar\rho_{ii\sigma}\geq\bar\rho_{jj\sigma}$, and
divide $\delta\rho(t)$ into particle-hole and hole-particle parts.
Introducing matrices in the ordered basis, so that $i<j$ and $m<n$
\begin{align}
 A_{ij\sigma,mn\sigma'}(\omega) &=
 \left[L_0+L_1^{r}(\omega)\right]_{ij\sigma,mn\sigma'}
 \\
 B_{ij\sigma,mn\sigma'}(\omega) &=
 \left[L_0+L_1^{r}(\omega)\right]_{ij\sigma,nm\sigma'}
\end{align}
and using general relations
\begin{align}
  \delta\rho_{ij\sigma}(\omega) &= \delta\rho^{*}_{ji\sigma}(-\omega)
 \\
  \left[L_0\right]_{ij\sigma,mn\sigma'} &= 
 -\left[L_0\right]_{ji\sigma,nm\sigma'}^{*}
 \\
  \left[L_1^{r}( \omega)\right]_{ij\sigma,mn\sigma'} &= 
 -\left[L_1^{r}(-\omega)\right]_{ji\sigma,nm\sigma'}^{*}
 \\
 \vec{\cal E}^r_{ij}(\omega) &= -\left[\vec{\cal E}^r_{ji}(-\omega)\right]^{*}
\end{align}
we get matrix equation of the form
\begin{align}
 \label{e_deltaGlt_reduced_FT}
 &\left[
 \begin{array}{cc}
  \omega - \mathbf{A}(\omega) & -\mathbf{B}(\omega) \\
  \mathbf{B}^{*}(-\omega) & \omega + \mathbf{A}^{*}(-\omega)
 \end{array}
 \right]
 \left[
 \begin{array}{c}
  \delta\mathbf{\rho}(\omega) \\ \delta\mathbf{\rho}^{*}(-\omega)
 \end{array}
 \right]
 \\
 &=\left[
 \begin{array}{c}
  \mathbf{S}^r(\omega) \\ -\left[\mathbf{S}^r(-\omega)\right]^{*}
 \end{array}
 \right]
 \nonumber
\end{align}
where
\begin{equation}
 \mathbf{S}^r(\omega) \equiv \vec{\mathbf{\cal E}}^r(\omega)\,\vec E(\omega)
\end{equation}
Eq.(\ref{e_deltaGlt_reduced_FT}) is reduced dimension form of 
Eq.(\ref{e_deltaGlt_final_FT}).
In the case when contacts are absent and real molecular orbitals are
chosen as a basis, matrices $\mathbf{M}=\{\mathbf{A},\mathbf{B},\mathbf{S}^r\}$
satisfy
\begin{equation}
 \mathbf{M}^{*}(-\omega) =\mathbf{M}(\omega)
\end{equation}
and Eq.(\ref{e_deltaGlt_reduced_FT}) reduces to the result derived within TDDFT
approach previously.\cite{TretiakChernyak2003}

Polarization of the molecule is
\begin{equation}
 P_\alpha(t) = \mbox{Tr}\left[\mu_\alpha\,\delta\rho(t)\right]
 = \sum_{\beta}\int_{-\infty}^{t}dt_1\,R^{(1)}_{\alpha\beta}(t-t_1)\,
   E_\beta(t_1) 
\end{equation}
where $\alpha,\beta=(x,y,z)$, $R^{(1)}_{\alpha\beta}(t-t_1)$ is the time-domain 
linear response function (tensor in $x,y,z$),
and trace is over molecular electronic basis 
($\mbox{Tr}\left[\ldots\right]=\sum_{i\sigma}[\ldots]_{ii\sigma}$).
Eq.(\ref{e_deltaGlt_final_FT}) yields for the
frequency-domain linear polarizability (FT of $R^{(1)}_{ij}(t-t_1)$)
\begin{equation}
 \label{alpha}
 \alpha_{\alpha\beta}(\omega) = 
 -\mbox{Tr}\left\{\mu_\alpha\left[\omega-L_0-L_1^r(\omega)\right]^{-1}
                 {\cal E}^r_\beta(\omega)\right\}
\end{equation}
Excitation energies (continuous spectrum of excitation energies) 
and oscillator strengths can be obtained from the poles and residues of
the polarizability. In the absence of contacts Eq.(\ref{alpha}) reduces
to the result derived previously.\cite{ChernyakMukamel2000}

\section{\label{conclude}Conclusion}
We discuss correspondence between NEGF and TDDFT
approaches, and derive approximate DM representation for NEGF equations.
Thus derived NEGF-TDDFT approach is superior over NEGF-DFT due to
ability to treat both time-dependent transport and/or
optical response of current-carrying molecular junctions.
However it partially misses nonlocality in time, inherent to NEGF.
Also fundamental question of combining the two ideologically different 
schemes (NEGF and TDDFT) remains open.
We further propose a practical scheme for calculation of linear optical
response of current carrying molecular junctions. Situation considered
corresponds to the case when an electron transport through the junction
is much faster than the characteristic time of an external laser field
(e.g. a pulse length).

First we derive expression for steady-state current
through the junction within NEGF-TDDFT scheme and discuss its
correspondence to the TDDFT-NEGF approaches 
to time-dependent transport of
Refs.~\cite{KoentoppBurkeCar,GebauerBurkeCar,GrossRubioLeeuwen,StefanucciAlmbladh_EL,StefanucciAlmbladh_PRB,StefanucciAlmbladhGross}. 
Formal equivalence of NEGF-TDDFT and TDDFT-NEGF schemes 
in the case of established steady-state flux permits to propose a way
to estimate $xc$ potential, and thus go beyond ground-state 
$xc$ potential implemented in the standard NEGF-DFT transport schemes.
 
After that we consider response of such current carrying junction
to weak external laser fields. We derive expressions for linear response
and polarizability, which in the absence of contacts 
reduce to previously obtained TDDFT results for optical response of 
isolated molecules. Presense of the contacts introduces memory 
in both Liouvillian and source term due to external field.  
Equation for response DM is then expressed in reduced (particle-hole and 
hole-particle) basis, which allows direct comparison to corresponding 
isolated molecule expression obtained previously.

The proposed approach allowing to calculate optical response of current 
carrying 
junctions is a first step in direction of ab initio calculations of such kind,
with the final goal to go beyond model based studies of optical properties of
current-carrying junctions available in the literature. 
Development of such ab initio schemes is especially important in 
light of experimental data on opto-electronic properties of molecular 
junctions which start to appear. In contrast to optical response of an isolated 
molecule presence of contacts will smear discrete excitation spectrum 
into continuous one, which is in complete analogy to smearing of
discrete energy spectrum of isolated molecule into continuous density of
states upon attaching to the contacts. Presence of contacts also allows for 
charge transfer transitions upon optical excitation -- process absent in the 
isolated molecule case, and thus changing the optical resonse of the 
molecule. Finally, nonequilibrium character of the junction will 
alternate optical response. An obvious change is presence of electronic
density in an extended energy region, defined by difference of 
electro-chemical potentials of the two contacts, opposed to 
well-defined ground state energy in the case of isolated molecule.   
For example, this may lead to appearance of additional (inverse) Raman 
scattering channel (situation when initial and final electronic state 
is LUMO rather than HOMO in the isolated molecule case), and to 
interference between the two (normal and inverse) channels, as is discussed
in detail elsewhere.\cite{Raman}

Consideration of situation, when time of electron
tunneling and characteristic time of external field change are comparable, 
is a subject of future studies.

\begin{acknowledgments}
M.G. thanks Alexander V. Balatsky for helpful discussions.
This contribution is supported by LANL LDRD program. M.G. gratefully
acknowledges the support of a LANL Director's Postdoctoral Fellowship. The
authors acknowledge support of Center for Integrated Nanotechnology (CINT)
and Center for Nonlinear Studies (CNLS) at Los Alamos National Laboratory
operated by Los Alamos National Security, LLC, for the National Nuclear
Security Administration of the U.S. Department of Energy under Contract No.
DE-AC52-06NA25396.
\end{acknowledgments}

\appendix
\section{\label{A}Approximate way to estimate $\bar v^{xc}$}
Here we discuss an approximate way to estimate $\bar v^{xc}$ 
for current-carrying molecular junction in steady-state.
We follow approach by van~Leeuwen\cite{Leeuwen1996}, where
connection between $xc$ potential and standard diagrammatic
technique was established. Ref.~\onlinecite{Leeuwen1996} 
considers a system both within full many-body approach and
within TDDFT, where true Hamiltonian $\hat H$ is replaced by 
fictitious single-particle Kohn-Sham Hamiltonian $\hat H^{(KS)}$ with 
some $xc$ potential $v^{xc}$ to be found.
Then adiabatic switching on of the interaction, $\hat H-\hat H^{(KS)}$,
is considered, which leads to connection between full many-body GF $G$
and single-particle Kohn-Sham GF $G_{KS}$ in the form
\begin{align}
 \label{G_GKS}
 &G(\vec r_1,\tau_1;\vec r_2,\tau_2) = G_{KS}(\vec r_1,\tau_1;\vec r_2,\tau_2)
 \nonumber \\
 &+ \int d\vec r_3 \int_c d\tau_3\int d\vec r_4\int_c d\tau_4 \,
   G_{KS}(\vec r_1,\tau_1;\vec r_3,\tau_3)
 \\ &\quad\times
 \left[
   \Sigma_{xc}(\vec r_3,\tau_3;\vec r_4,\tau_4)-
   \delta(\vec r_3-\vec r_4)\delta(\tau_3,\tau_4)v^{(xc)}(\vec r_3,\tau_3)
   \right]
 \nonumber \\ &\quad\times
 G(\vec r_4,\tau_4;\vec r_2,\tau_2)
 \nonumber
\end{align}
with $\Sigma_{xc}$ being $xc$ part of the SE due to electron-electron 
interaction. Since the two approaches are assumed to give the same 
electron density
\begin{equation}
 n(\vec r,t) = -iG^{<}(\vec r,t;\vec r,t) = -i G_{KS}^{<}(\vec r,t;\vec r,t)
\end{equation}
lesser projection of (\ref{G_GKS}) provides a (self-consistent) 
expression for $xc$ potential $v^{(xc)}$ in terms of SE $\Sigma_{xc}$
\begin{align}
 \label{vxc_Sigmaxc}
 &\int d\vec r_1 \int_{-\infty}^{+\infty} dt_1\,
 \left[G_{KS}(\vec r,t;\vec r_1,t_1)\,v^{xc}(\vec r_1,t_1)\,
       G(\vec r_1,t_1;\vec r,t)\right]^{<}
 \nonumber \\ &
 =\int d\vec r_1\int_{-\infty}^{+\infty} dt_1
 \int d\vec r_2\int_{-\infty}^{+\infty} dt_2\,
 \\
 &\qquad\left[G_{KS}(\vec r,t;\vec r_1,t_1)\,\Sigma_{xc}(\vec r_1,t_1;\vec r_2,t_2)\,
       G(\vec r_2,t_2;\vec r,t)\right]^{<} 
 \nonumber
\end{align}
where
\begin{align}
 [G_{KS}\, v^{(xc)}\,G]^{<}   &= G_{KS}^{<}\, v^{(xc)}\, G^{a}
                               + G_{KS}^{r}\, v^{(xc)}\, G^{<}
 \\
 [G_{KS}\,\Sigma_{xc}\,G]^{<} &= G_{KS}^{<}\,\Sigma_{xc}^{a}\, G^{a}
                               + G_{KS}^{r}\,\Sigma_{xc}^{<}\, G^{a}
 \\
                              &+ G_{KS}^{r}\,\Sigma_{xc}^{r}\, G^{<}
 \nonumber
\end{align}
Eq.(\ref{vxc_Sigmaxc}) is our starting point. 

Now we consider situation of established steady-state current in the system. 
In this case it was shown that $v^{xc}$ becomes constant far from the
device region.\cite{StefanucciAlmbladh_PRB}
Accordingly, we formally partition the system into 3 parts: two contacts 
($K=L,R$), where $v^{xc}$ is constant, and device (molecule, $M$),
where it varies in space. Then integral over space in the left of 
Eq.(\ref{vxc_Sigmaxc}) splits into two parts (integration over
contacts $V_K$ and over device $V_M$), while double space integral
in the right will have 4 contributions: two over $V_K$ and $V_M$,
respectively,
and two mixed ones. The last may be neglected (especially when device
is well separated from the contacts) taking into account relatively
short range of $xc$ interaction. This approximation yields additive 
(in $K$ and $M$) structure of Eq.(\ref{vxc_Sigmaxc}), and obviously
one can equate corresponding parts separately.
As a result one gets for the device region only expression similar to
Eq.(\ref{vxc_Sigmaxc}), where $v^{(xc)}$ is time-independent and where
GFs are restricted to the device region only. As was demonstrated
in Ref.~\onlinecite{StefanucciAlmbladh_PRB} the last are given formally
from the usual NEGF expressions for GFs of the device region
in terms of SEs due to coupling to the contacts, when appropriate
constant shift of chemical potentials in the contacts due to
xc response is taken into account.  
Moreover, while generally this shift may depend on the history
of appearance of the steady-state, in LDA it depends only on the
instantaneous local density and has no memory at 
all,\cite{StefanucciAlmbladh_PRB} i.e. instantaneous (and constant in time)
potential and instantaneous (and constant in time) density uniquely define each 
other. In this case we can assume that the correct (shifted by $xc$ response)
{\em chemical potentials in the contacts are some predefined boundary 
conditions},
as is usually done within NEGF, which determine situation in the
device region. In other words we assume equivalence of partitioned NEGF
and partitionless TDDFT approaches in this case. Then, after introducing
some basis within the device region (so that GFs, SEs, and $v^{xc}$ are
matrices in Hilbert space), Eq.(\ref{vxc_Sigmaxc}) yields
\begin{align}
 \label{ss_vxc_Sigmaxc}
 &\int\frac{dE}{2\pi}\,\left[G_{KS}(E)\, \bar v^{xc}\, G(E)\right]^{<} =
 \\
 &\int\frac{dE}{2\pi}\,\left[G_{KS}(E)\,\Sigma_{xc}(E)\, G(E)\right]^{<}
 \nonumber
\end{align}
with GFs given by usual NEGF expressions. Eq.(\ref{ss_vxc_Sigmaxc})
is an approximate way to estimate $\bar v^{xc}$
in the steady-state situation. Explicit expression for $\bar v^{xc}$ is 
obtained by rewriting Eq.(\ref{ss_vxc_Sigmaxc}) in Liouville space.

\end{document}